\documentclass[letter,
               keeplastbox,   
               ]{jacow}
\usepackage{pdfpages,multirow,ragged2e,subcaption,verbatim} %
\makeatletter%
	\ifboolexpr{bool{xetex}}
	 {\renewcommand{\Gin@extensions}{.pdf,%
	                    .png,.jpg,.bmp,.pict,.tif,.psd,.mac,.sga,.tga,.gif,%
	                    .eps,.ps,%
	                    }}{}
\makeatother

\ifboolexpr{bool{xetex} or bool{luatex}} 
 {}                                      
 {\usepackage[utf8]{inputenc}}           

\usepackage[USenglish]{babel}

\graphicspath{{./}{./figures}}
\ifboolexpr{bool{jacowbiblatex}}%
 {%
  \addbibresource{jacow-test.bib}
  \addbibresource{biblatex-examples.bib}
 }{}
\newcommand{\pyorbit}{{\tt PyORBIT}}

\begin{document}

\title{Full-Cycle Simulations of The Fermilab Booster \thanks{This manuscript has been authored by Fermi Research Alliance, LLC under Contract No. DE-AC02-07CH11359 with the U.S. Department of Energy, Office of Science, Office of High Energy Physics.}}

\author{J.-F. Ostiguy \thanks{ostiguy@fnal.gov}, C. M. Bhat \\ Fermilab, Batavia, IL, USA}
	
\maketitle

\begin{abstract}
The Proton Improvement Plan phase II (PIP-II) project currently under construction at FNAL will replace the existing
400 MeV normal conducting linac with a new 800 MeV superconducting linac.
The beam power in the downstream rapid-cycling Booster synchrotron will be
doubled by raising the machine cycle frequency from 15 to 20 Hz and by increasing the
injected beam intensity by a factor 1.5. This has to be accomplished without
raising uncontrolled losses beyond the administrative limit of
500 W. In addition, slip-stacking efficiency in the Recycler,
the next machine in the accelerator chain, sets an upper limit on
the longitudinal emittance of the beam delivered by the Booster. 
As part of an effort to better understand potential losses and emittance blow-up
in the Booster, we have been conducting full cycle 6D simulations using the code \pyorbit.
The simulations include space charge, wall impedance effects and transition   
crossing. In this paper, we discuss our experience with the code and present
representative results for possible operational scenarios.
 \end{abstract}
\section{INTRODUCTION}

The Proton Improvement Plan phase II (PIP-II) will double the beam power available for neutrino production in support of the laboratory flagship DUNE experiment.  In addition to a new superconducting linac, modifications to the existing complex will be needed. For safe operation and on-hands maintenance, a rule of thumb is that uncontrolled particle losses in circular machines in the accelerator chain should not exceed 1 W/m. Since the Booster already operates near this limit, an increase in beam power during the PIP-II era implies a reduction in particle losses while preserving beam stability and limiting emittance blowup. Therefore, we have been conducting full 6D particle tracking simulations in combination with extensive experimental investigations. In this paper we focus primarily on excessive longitudinal emittance blowup at transition, an issue of concern for the Booster. During injection, both the transverse and longitudinal distributions will be painted in phase space to enhance the charge spatial density uniformity, thereby reducing space charge tune shift and peak longitudinal line density. At the start of the acceleration cycle, the painted emittance is 0.06 eV-s (95\%).  To prevent  excessive losses during slip-stacking at injection in the Recycler ring, the longitudinal emittance of the beam transferred out from the Booster should not exceed 0.1 eV-s (95\%). 
\section{BOOSTER}
 The PIP-II era Booster will accelerate protons from 0.8 to 8 GeV. Its lattice comprises 24 cells with basic structure OFDOODF, where F and D respectively represent focusing and defocusing gradient bending magnets. The acceleration voltage is provided by a total of 22 cavities. To prevent multipactoring, they are operated at full voltage in two paraphased groups; the net ring voltage is controlled by altering the phase between the two groups. The maximum total available ring voltage is 1.16 MV.

 The current injection scheme is based on charge-exchange multi-turn injection followed by rf adiabatic capture. In contrast, PIP-II will transfer linac bunches directly into already
formed stationary rf buckets. Because the linac and ring frequencies are not commensurable, bunches that would not land inside ring rf buckets will be removed
in the MEBT using a fast bunch-by-bunch kicker. Injection into a stationary bucket allows for transverse and longitudinal phase space painting to reduce the incoherent tune shift and the peak longitudinal density by making the charge spatial density more uniform.  

A distintive feature of the Fermilab Booster is the absence of a metallic vacuum chamber; instead, the entire magnet aperture region is under vacuum.  This approach side-steps issues with magnetic shielding and magnetic field perturbations caused by eddy currents induced in the chamber wall. A downside  is that the beam is directly exposed to the magnet laminations resulting in an unusually lossy wall impedance. This impedance causes significant distortion of the phase space at transition.

The Booster crosses transition at  $\gamma_t\simeq5.45$. The non-adiabatic dynamics at transition results in an increase in longitudinal emittance; due to space charge and impedance, a portion of this increase is intensity dependent. Keeping losses at an acceptable level during slip-stacking in the Recycler ring imposes a limit of 0.1 eV-s on the final 95\% longitudinal emittance.  
Currently, there is no dedicated $\gamma_t$ jump system; however emittance blowup can be mitigated using an active quadrupole mode damper. The Booster is equipped with 48 quadrupole corrector magnets and investigations are ongoing to determine to what extent these could be used to emulate a $\gamma_t$ jump system. 
\section{SIMULATIONS }
Although a number tracking codes capable of handling space charge have been developed, few of them are adequate to simulate a rapid-cycling synchrotron. In addition, many codes were developed decades ago and do not support parallel execution. Other important considerations are the existence of an active user community and the availability of the source code under non-restrictive terms. All these factors led us to select \pyorbit, a code developed at ORNL \cite{PYORBIT} originally to model the the SNS accumulation ring. While \pyorbit\  supports a variety of space charge solvers, the bunch longitudinal to transverse size ratio in the Booster is  high $(\gg 10)$ making it appropriate to employ a so-called 2.5D solver where the transverse and longitudinal space charge forces are computed separately. Longitudinally, the force is obtained from the derivative of the longitudinal line density while transversely, it is obtained by solving Poisson's equation in the beam frame using FFTs and weighting the resultant field pattern by the local longitudinal line density. The 2.5D approximation dramatically reduces computational load. To fix ideas, on a modern small computer cluster, simulating a typical full 25 ms booster acceleration cycle (15000 turns) involving 50 to 100k particles) on 20 CPU cores typically takes 6 hours.     
\begin{figure*}[!h]
\centering
\includegraphics[width=0.95\textwidth,height=0.5\textwidth ]{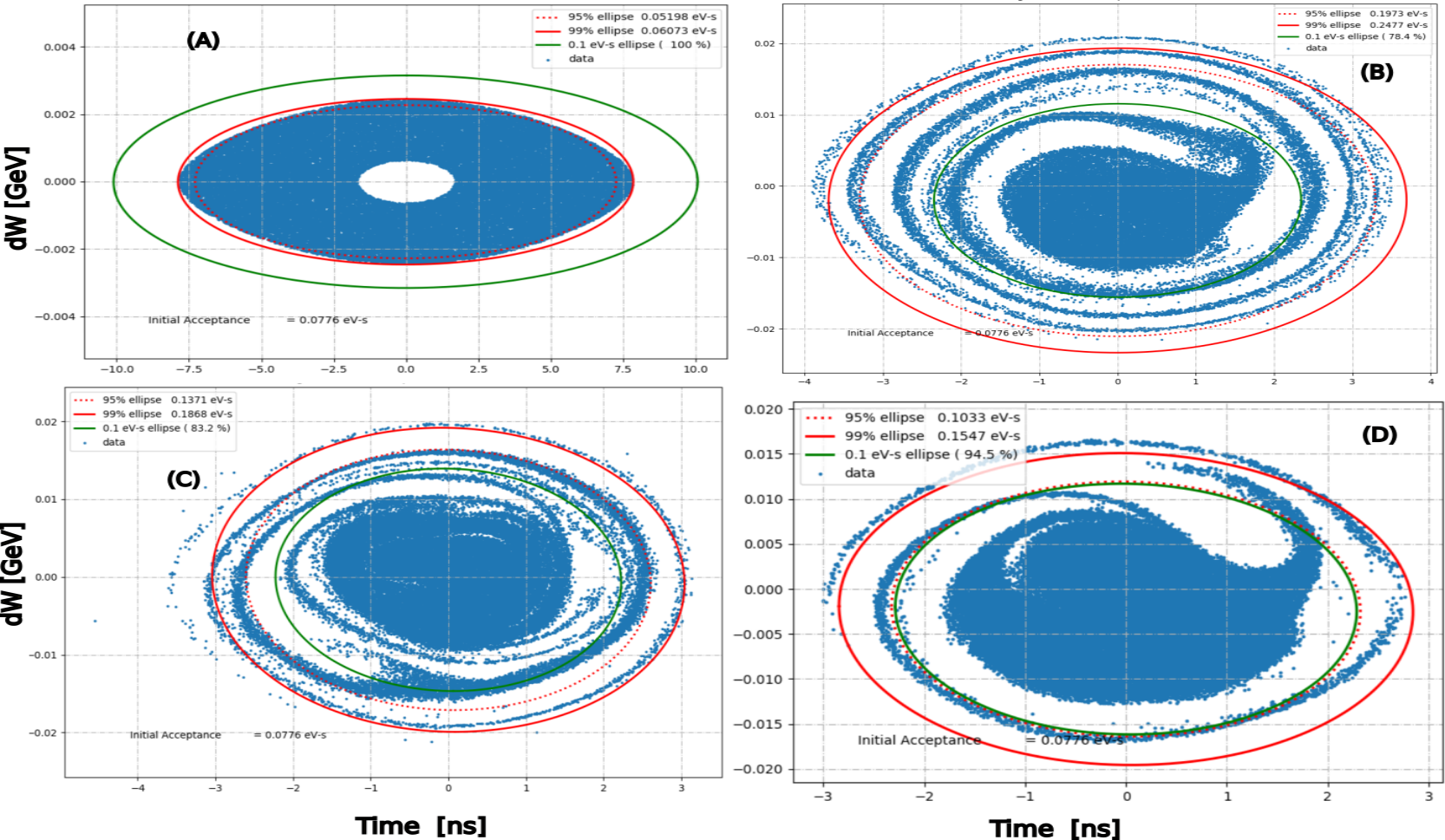} 
\caption{Longitudinal phase space distributions of a bunch at injection and extraction.}
\label{fig:THPC69_f1}
\end{figure*}
%

We encountered a number of issues when attempting to simulate a full cycle through transition. In contrast with a longitudinal only code,
the precise value of $\gamma_t$, the dependence of the the momentum compaction factor on $\Delta p/p$ are not direct input parameters; rather, they are implicitly determined by the lattice. Since the precise value of $\gamma_t$ associated with a lattice model can be affected by the element slicing scheme, we found that the best approach is to have \pyorbit\ itself compute $\gamma_t$.
While this is possible, the linear optics functionality of the code has limitations that often need
to be circumvented. 
Just like in real machine, unless the lattice transverse tunes are carefully set a priori, the change in $\Delta p/p$ at transition can easily excite transverse resonances causing rapid beam loss. While physically obvious, this type of issue can be difficult to diagnose numerically.

As already mentioned, the Booster wall impedance is unsually lossy. Near transition, the bunch becomes narrow, the voltage induced by the wall reaches a maximum and the resistive part of the impedance causes energy loss. Like many tracking codes, \pyorbit\ particle coordinates are expressed with respect to an a-priori defined reference, the so-called synchronous particle. Given an rf voltage  amplitude curve $V(t)$, the rf phase $\phi_s(t)$ of this reference particle is set by condition  $V(t) \sin \phi_s(t) = V_\text{sync}(t)$ where $V_\text{sync}$ is the voltage required to preserve synchronism.

Due to the lossy impedance, to remain in synchronism the bunch centroid will move adiabatically 
to a phase $\phi_{s,\text{bunch}}$. In the Booster, near transition, $\phi_{s,\text{bunch} }- \phi_s$ is on the order of $10$ deg.
To preserve the dynamics across transition, the bunch phase must jump from $\phi_{s,\text{bunch}}$ to $\pi-\phi_{s,\text{bunch}}$. It is important to appreciate that in the presence of a lossy impedance, the bunch centroid phase and the synchronous phase $\phi_s$ will not be the same. As a result, simply jumping the rf phase from $\phi_{s}$ to $\pi-\phi_s$ would cause spurious mismatch. In the real machine, there is no notion of the reference phase $\phi_s$; the beam centroid (with respect to the rf phase) is simply phase locked to the rf frequency. At transition both the rf phase signal and the beam phase signal are inverted resulting in a net change $\pi -2 \phi_{s,\text{bunch}}$. 
There are different ways to address this issue in simulation. Our approach mimics the phase lock mechanism. At each turn, the bunch centroid phase and the reference phase are compared; both the rf phase and the longitudinal particle coordinates are then shifted so that the coordinate of the particles are always expressed with respect to the bunch centroid. 
Investigating longitudinal emittance blowup required the inclusion of a quadrupole damper in the simulations.  While \pyorbit\ does not readily provide direct support for dampers or feedback systems, it is a relatively straightforward matter to implement such functionality at the {\tt python} level. One can show \cite{DAMPERMODEL} that in the linear approximation, the second moment of the bunch longitudinal 
coordinate obeys the equation
\begin{equation}
\frac{d^2}{dt^2} \Delta \sigma^2_\phi + 4 \omega_S^2 \Delta \sigma^2_\phi = -2 \omega_S\sigma^2_0 \epsilon(t)
\label{e2}
\end{equation}
where  $\omega_S$ is the synchrotron frequency,  $\sigma_0^2$ is the matched (equilibrium) bunch length, $\Delta \sigma^2_\phi =  \sigma^2_\phi- \sigma^2_0$ and  $\epsilon(t)=\frac{\Delta V}{V}$ is an rf voltage modulation.
If one uses as modulation a signal proportional to the derivative of the second moment  
\begin{equation}
  \epsilon(t)= \frac{2\xi}{\sigma_0 \omega_S} \frac{d}{dt}\Delta \sigma^2_\phi
\label{e3}
\end{equation}
where $ 0 < \xi  <1$, Eq. \eqref{e3} becomes 
\begin{equation}
\frac{d^2}{dt^2} \Delta \sigma^2_\phi + 4 \omega_S \frac{d}{dt} \Delta \sigma^2_\phi + 4 \omega_S^2 \Delta \sigma^2_\phi = 0 
\label{e4}
\end{equation}
This is a standard second order ODE; its solution is a damped oscillation with damping time $\tau = \frac{1}{2\xi\omega_S}$.   
In the code the damper is implemented by taking a properly scaled and filtered version of the derivative of the envelope of the second moment signal to modulate the rf cavity voltage. 
\section{RESULTS}
Figure \ref{fig:THPC69_f1} (a) shows the initial longitudinal phase space at the start of the acceleration cycle. The hollowed out region is created during injection painting to 
reduce the peak line density. The 95\% emittance is 0.052 eV-s.  For comparison purposes, we first present Fig. \ref{fig:THPC69_f1}(b) which shows the phase space near the end of the acceleration cycle assuming only a simple phase jump at transition. The final 95\% emittance is 0.197 eV-s, about a factor 2 larger than the requirement. 

\begin{figure}[!h]
\centering
\includegraphics[width=0.95\columnwidth,height=0.55\columnwidth ]{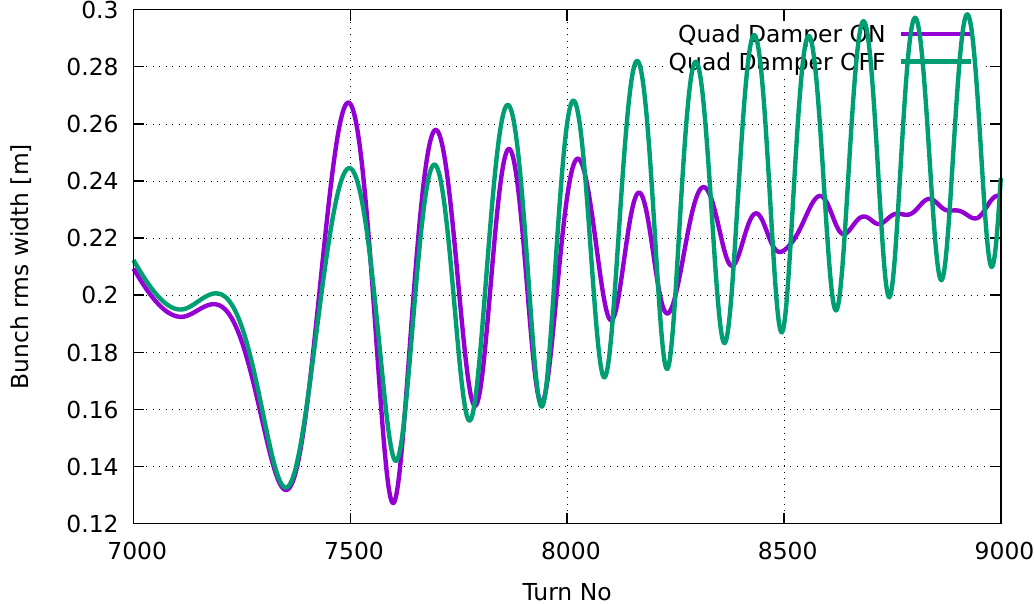}
\caption{Damping quadrupole oscillations.} 
\label{fig:THPC69_f2}
\end{figure}

\begin{figure}[!h]
\centering
\includegraphics[width=0.95\columnwidth,height=0.4\columnwidth]{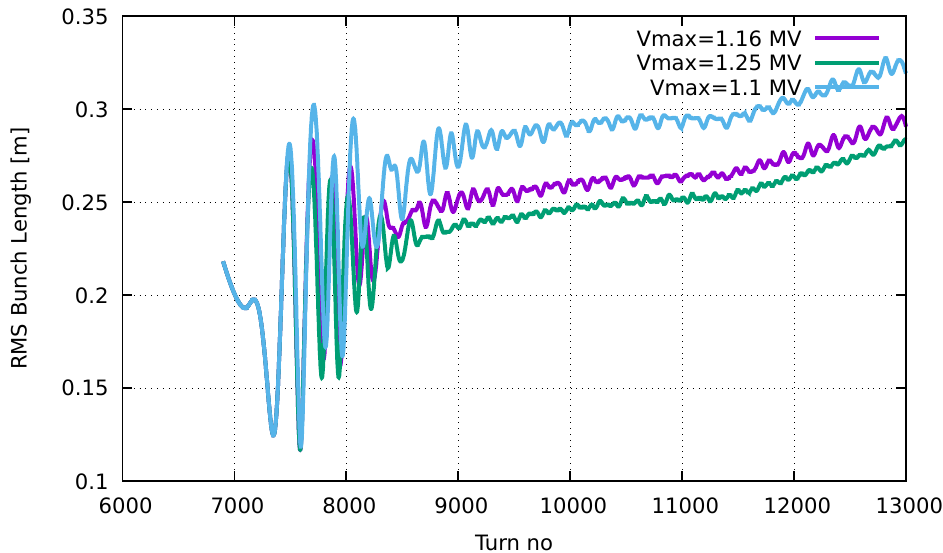}
\caption{Voltage overhead and damper performance.} 
\label{fig:THPC69_f3}
\end{figure}

Fig. \ref{fig:THPC69_f2} shows the rms bunch length oscillations triggered by crossing transition ( transition occurs at turn 7100). The green trace correspond to the case where the damper is turned off. The magenta trace correspond to the case where the damper is turned on, showing that a damper is  clearly effective in reducing the amplitude of the bunch shape oscillations and preventing the phase space filamentation responsible for the emittance blowup. The phase space at the end of a cycle where the damper is turned on is shown in Fig. \ref{fig:THPC69_f1}(c); the 95\% emittance has been reduced to $0.137$ eV-s. An issue with the damper is that the amplitude of the modulation is limited by the available ring voltage; this is amounts to a feedback gain saturation effect.
The availability of a modest voltage overhead can significantly affect the damper effectiveness. Fig. \ref{fig:THPC69_f3} illustrates this effect. In the simulation, the maximum available voltage (initially set to 1.1 MV) is increased to 1.16 (cyan) and 1.25 (green) MV;the final bunch length is (therefore the final emittance) can clearly be further reduced if additional voltage is available.       

An option under consideration is to emulate the old system by pulsing the existing quadrupole correctors (there is a focusing and a defocusing corrector in each of the lattice 24 cells). By exciting a subset of the correctors in a spatial pattern that excites a closed orbit resonance, it appears feasible to create enough of a perturbation to change $\gamma_t$ somewhere between 0.3 and 0.6 units.  How much is achievable in practice depends on a number of factors that are still under study. Another issue is the corrector power supply slew rate limit.  As an example, Fig. \ref{fig:THPC69_f1}(d) shows the final phase space distribution obtained using such an emulated $\gamma_t$ jump while the quadrupole damper is turned off. In this particular example $\Delta \gamma_t \sim 0.6 $ and the jump duration is $\sim 100 \mu$s (somewhat shorter than the Booster non-adiabatic time -- about 210 $\mu{\rm s}$ at 20 Hz). In this case, it is seen that  the final 95 \% emittance ($0.103$ eV-s) would meet the PIP-II requirement.
\section{CONCLUSION}
We demonstrated full cycle tracking simulations using the code \pyorbit. Albeit based
on a simplified model, the simulation predicts that using the existing quadrupole damper, the Booster longitudinal emittance would come slightly above the PIP-II requirement. To further reduce emittance blowup,  a possible scenario would be to combine the quadrupole damper with a weak $\gamma_t$ jump based on pulsing existing correctors.  Both experimental and simulation work are on-going to determine the viability of this approach.

%
%
\ifboolexpr{bool{jacowbiblatex}}%
	{\printbibliography}%

\begin{thebibliography}{9} 
	\bibitem{PYORBIT}
		A. Shislo \textit{et al}., The particle accelerator simulation code PyORBIT, Procedia Comput. Sci., vol 51, 2015, pp. 1272-1281\\   \url{doi:10.1016/j.procs.2015.05.312} 
	\bibitem{DAMPERMODEL}
		H. Klingbeil \textit{et al.}, Modeling longitudinal oscillations of bunched beams in synchrotrons\\ \url{doi:10.48550/arXiv.101.3957}
	\end{thebibliography}
	{%
	
	
} 

%
%


\end{document}